\documentclass{jpp}

\usepackage{graphicx}
\usepackage{natbib}

\usepackage{graphics}
\usepackage{epsfig}
\usepackage{bm}
\usepackage{amsmath,amssymb}
\usepackage{stmaryrd}

\def\ov#1{\overline{#1}}

\def\wt#1{\widetilde{#1}}
\def\vb#1{\mbox{\boldmath$#1$}}
\def\pd#1#2{\frac{\partial #1}{\partial #2}}

\def\wh#1{\widehat{#1}}
\def\bdot{\,\vb{\cdot}\,}
\def\btimes{\,\vb{\times}\,}

\def\bhat{\wh{{\sf b}}}
\def\cal#1{\mathcal{#1}}

\def\exd{{\sf d}}
\def\bhat{\wh{{\sf b}}}

\newcommand{\bc}{\begin{center}}
\newcommand{\ec}{\end{center}}
\newcommand{\bt}{\begin{tabbing}}
\newcommand{\et}{\end{tabbing}}
\newcommand{\be}{\begin{equation}}
\newcommand{\ee}{\end{equation}}
\newcommand{\ba}{\begin{eqnarray}}
\newcommand{\ea}{\end{eqnarray}}

\title[Polarization Effects in Higher-order Guiding-center Lagrangian Dynamics]{Polarization Effects in Higher-order Guiding-center Lagrangian Dynamics}

\author[Brizard]%
{Alain J.~Brizard%
  \thanks{Email address for correspondence: abrizard@smcvt.edu}}

\affiliation{Department of Physics, Saint Michael's College, Colchester, VT 05439, USA}

\date{?; revised ?; accepted ?. - To be entered by editorial office}
\begin{document}

\maketitle

\begin{abstract}
The extended guiding-center Lagrangian equations of motion are derived by Lie-transform perturbation method under the assumption of time-dependent and inhomogeneous electric and magnetic fields that satisfy the standard guiding-center space-time orderings. Polarization effects are introduced into the Lagrangian dynamics by the inclusion of the polarization drift velocity in the guiding-center velocity and the appearance of finite-Larmor-radius corrections in the guiding-center Hamiltonian and guiding-center Poisson bracket.
\end{abstract}

\section{Introduction}

Polarization effects have a rich history in plasma physics \citep{Pfirsch_1984,Pfirsch_Morrison_1985,Cary_Brizard_2009}. Their importance stems from the assumption of quasineutrality in a strongly magnetized plasma and the dielectric properties of a guiding-center plasma \citep{Hinton_1984}. While these effects are traditionally associated with the presence of an electric field in a magnetized plasma \citep{Itoh_1996,Hazeltine_Meiss_2003,Wang_Hahm_2009,Joseph_2021,Brizard_2023}, they are also associated with magnetic drifts \citep{Kaufman_1986,Brizard_2013,Tronko_Brizard_2015}.

Recently, second-order terms in guiding-center Hamiltonian theory (in the absence of an electric field) were shown to be crucial \citep{Brizard_Hodgeman_2023} in assessing the validity of the guiding-center representation in determining whether guiding-center orbits were numerically faithful to the particle orbits in axisymmetric magnetic geometries, which partially confirmed earlier numerical studies in axisymmetric tokamak plasmas \citep{Belova_2003}. In particular, it was shown that a second-order correction associated with guiding-center polarization \citep{Kaufman_1986,Brizard_2013,Tronko_Brizard_2015} was needed in order to obtain faithful guiding-center orbits. 

Indeed, without the inclusion of second-order effects, it was shown that, within a few bounce periods after leaving the same physical point in particle phase space, a first-order guiding-center orbit deviated noticeably from its associated particle orbit, while a second-order guiding-center orbit followed the particle orbit to a high degree of precision \citep{Brizard_Hodgeman_2023}. In addition, as initially reported by \cite{Belova_2003}, the guiding-center Hamiltonian formulation  is a faithful representation of the particle toroidal angular momentum \citep{Tronko_Brizard_2015,Brizard_Hodgeman_2023}, which is an exact particle constant of motion in an axisymmetric magnetic field, only if second-order effects are included.

\subsection{Lagrangian dynamics in extended phase space}

In the present work, we consider time-dependent and inhomogeneous electric and magnetic fields (which still satisfy the guiding-center space-time orderings $|\nabla|^{-1} \gg \rho_{\rm th} = v_{\rm th}/\Omega$ and $\partial/\partial t \ll \Omega = eB/mc$) and we assume that the $E\times B$ velocity ${\bf u}_{\rm E} = {\bf E}\btimes c\bhat/B$ is comparable to the particle's thermal velocity $v_{\rm th}$. Because of the explicit time dependence of the electromagnetic fields, the Lagrangian charged-particle dynamics takes place in an odd-dimensional space $({\bf q},{\bf p},t)$, where the non-autonomous Hamiltonian $H({\bf q},{\bf p},t)$ is a function of the canonical coordinates $({\bf q},{\bf p})$, from which the canonical Hamilton equations $d{\bf q}/dt = \partial H/\partial{\bf p}$ and $d{\bf p}/dt = -\,\partial H/\partial{\bf q}$ are derived, as well as time $t$, from which we obtain the energy equation $dH/dt = \partial H/\partial t$ (i.e., energy is not conserved). 

The use of an extended phase space is a well-known method in classical mechanics \citep{Lanczos_1970} used to deal with a time-dependent Hamiltonian system by transforming it into an autonomous Hamiltonian system in an even-dimensional symplectic setting. Here, the canonical time-energy coordinates $(t,w)$ are included in the extended phase-space coordinates $({\bf q},t;{\bf p},w)$, where the space-time coordinates $({\bf q},t)$ are canonically conjugate to the momentum-energy coordinates $({\bf p},w)$, with the extended Hamilton equations $dw/ds = \partial{\cal H}/\partial t = \partial H/\partial t$ and $dt/ds = -\,\partial{\cal H}/\partial w = 1$, where ${\cal H} \equiv H({\bf q},{\bf p},t) - w$ is the extended Hamiltonian and a particle orbit in extended phase space (parametrized by $s$) takes place on the energy surface ${\cal H} = 0$, i.e., $w = H({\bf q},{\bf p},t)$.

Using the dimensional ordering parameter $\epsilon$ associated with the renormalized particle mass $m \rightarrow \epsilon\,m$ \citep{Brizard_1995}, instead of the standard ordering $e \rightarrow e/\epsilon$ \citep{Kulsrud_1983,RGL_1983}, we begin with the extended phase-space particle Lagrangian one-form
\begin{equation}
\gamma \;=\; \left( \frac{e}{c}\;{\bf A} \;+\; \epsilon\,{\bf p}_{0}\right)\bdot\exd{\bf x} \;-\; w\;\exd t \;\equiv\; \gamma_{0} \;+\; \epsilon\,\gamma_{1},
\label{eq:Gamma_particle}
\end{equation}
and the extended particle Hamiltonian
\begin{equation}
{\cal H} \;=\; e\,\Phi \;-\; w \;+\; \epsilon\,|{\bf p}_{0}|^{2}/2m \;\equiv\; {\cal H}_{0} + \epsilon\,{\cal H}_{1},
\label{eq:Ham_particle}
\end{equation}
where ${\bf p}_{0}$ denotes the local particle kinetic momentum at position ${\bf x}$. In the present work, we consider the standard ordering \citep{Kulsrud_1983} for the parallel electric field: ${\bf E} = {\bf E}_{\bot} + \epsilon\,E_{\|}\,\bhat$. In contrast to \cite{Madsen_2010} and \cite{Frei_2020}, who used the same mass ordering ($m \rightarrow \epsilon\,m$), we use extended (eight-dimensional) phase space in Eqs.~\eqref{eq:Gamma_particle}-\eqref{eq:Ham_particle}, where the energy coordinate $w$ is canonically conjugate to time $t$ \citep{RGL_1981,Cary_Brizard_2009}. This extended phase-space formulation yields a simple form for the extended Poisson bracket [see Eq.~\eqref{eq:gcPB_ext}], also adopted (without derivation) by \cite{Madsen_2010}, which plays an important role in the variational formulation of the guiding-center Vlasov-Maxwell equations \citep{Brizard_2023_gcVM}.

Here, the electric-field ordering implies that the local particle momentum 
\begin{equation}
{\bf p}_{0} \;\equiv\; p_{\|0}\,\bhat({\bf x},t) \;+\; {\bf P}_{\rm E}({\bf x},t) \;+\; {\bf q}_{\bot0}(J_{0},\theta_{0};{\bf x},t)
\label{eq:p_local}
\end{equation}
is decomposed into the gyroangle-independent parallel component $p_{\|0} \equiv {\bf p}_{0}\bdot\bhat$ and the $E\times B$ momentum
\begin{equation}
{\bf P}_{\rm E} \;\equiv\; {\bf E}\btimes\frac{e\bhat}{\Omega} \;=\; m\,{\bf u}_{\rm E}, 
\label{eq:P_E_def}
\end{equation}
and the gyroangle-dependent perpendicular momentum ${\bf q}_{\bot0} \equiv |{\bf q}_{\bot0}|\,\wh{\bot}$, respectively, where $J_{0} \equiv |{\bf q}_{\bot0}|^{2}/(2m\Omega)$ represents the lowest-order gyroaction and the gyroangle derivative 
$\partial{\bf q}_{\bot0}/\partial\theta_{0} \equiv {\bf q}_{\bot0}\btimes\bhat = -\,|{\bf q}_{\bot0}|\,\wh{\rho}$ introduces the rotating orthogonal unit-vector fields $(\bhat,\wh{\bot},\wh{\rho})$.

\subsection{Purpose of the present work}

The purpose of the present work is motivated by the need to derive higher-order guiding-center equations that can accurately describe the magnetic confinement of charged particles in regions with steep gradients (e.g., the pedestal region of advanced tokamak plasmas). For many situations of practical interest, the presence of a strong electric field is associated with strong plasma flows with steep sheared rotation profiles for which second-order effects (including finite-Larmor-radius effects) must be included in a self-consistent guiding-center theory \citep{Hahm_1996,Chang_2004,Lanthaler_2019,Frei_2020}. 

Guiding-center equations of motion with second-order corrections in the presence of time-independent electric and magnetic fields were derived using Lie-transform perturbation method by \cite{Brizard_1995} and \cite{Hahm_1996}, following the earlier work of \cite{RGL_1981}. These perturbation methods were also used by \cite{Miyato_Scott_2009}, \cite{Madsen_2010}, and \cite{Frei_2020}, who derived self-consistent guiding-center Vlasov-Maxwell equations that included guiding-center polarization and magnetization effects. Not all second-order effects were included in these models, however, and it is the purpose of the present work to derive a more complete higher-order guiding-center Vlasov-Maxwell theory, with a full representation of guiding-center polarization that can be directly derived by the guiding-center push-forward method \citep{Brizard_2013,Tronko_Brizard_2015}.

\section{\label{sec:Lie}Guiding-center Lie-transform Perturbation Analysis}

The derivation of the guiding-center equations of motion by Lie-transform perturbation method is based on a phase-space transformation from the (local) particle extended (eight-dimensional) phase-space coordinates $z_{0}^{\alpha} = ({\bf x}, p_{\|0}; J_{0},\theta_{0};w_{0},t)$, where the energy-time canonical coordinates $(w_{0},t)$ are included, to the guiding-center phase-space coordinates $Z^{\alpha} = ({\bf X}, P_{\|}; J, \theta;W,t)$ generated by the vector fields $({\sf G}_{1}, {\sf G}_{2}, \cdots)$:
\begin{equation}
Z^{\alpha} \;=\; z_{0}^{\alpha} \;+\; \epsilon\,G_{1}^{\alpha} \;+\; \epsilon^{2}\,\left( G_{2}^{\alpha} + \frac{1}{2}\,{\sf G}_{1}\cdot\exd G_{1}^{\alpha}\right) \;+\; \cdots,
\label{eq:z_bar_z} 
\end{equation}
and its inverse
\begin{equation}
z_{0}^{\alpha} \;=\; Z^{\alpha} \;-\; \epsilon\,G_{1}^{\alpha} \;-\; \epsilon^{2}\,\left( G_{2}^{\alpha} - \frac{1}{2}\,{\sf G}_{1}\cdot\exd 
G_{1}^{\alpha}\right) \;+\; \cdots.
\label{eq:zz_bar} 
\end{equation} 
In order for the particle time to be invariant under the guiding-center transformation \eqref{eq:z_bar_z}, we require that $G_{n}^{t} \equiv 0$ to all orders $n \geq 1$, i.e., the guiding-center time is identical to the particle time. From these generating vectors fields, the pull-back and push-forward Lie-transform operators ${\sf T}_{\rm gc} = \exp(\epsilon\pounds_{1})\,\exp(\epsilon^{2}\pounds_{2})\cdots$ and ${\sf T}_{\rm gc}^{-1} = \cdots \exp(-\epsilon^{2}\pounds_{2})\,\exp(-\epsilon\pounds_{1})$ are constructed in terms of Lie derivatives $(\pounds_{1}, \pounds_{2}, \cdots)$ generated by the vector fields $({\sf G}_{1}, {\sf G}_{2}, \cdots)$. More details about the Lie-transform perturbation method used in guiding-center theory can be found in \cite{RGL_1982} (as well as the unpublished UCLA report {\it Variational Principles for Guiding Center Motion} (PPG-611) written by Littlejohn in 1982), while the notation used here is taken from \cite{Tronko_Brizard_2015} and \cite{Brizard_Tronko_arxiv_2016}.

We now wish to derive the extended guiding-center phase-space Lagrangian one-form
\begin{equation}
\Gamma_{\rm gc} \;\equiv\; {\sf T}_{\rm gc}^{-1}\gamma + \exd \sigma = \Gamma_{0{\rm gc}} + \epsilon\,\Gamma_{1{\rm gc}} + \epsilon^{2}\,\Gamma_{2{\rm gc}} + \cdots,
\label{eq:ovgamma_Lie}
\end{equation}
where the gauge scalar field $\sigma = \epsilon\,\sigma_{1} + \epsilon^{2}\sigma_{2} + \cdots$ is chosen at each order in order to simplify the transformation, with
\begin{eqnarray}
\Gamma_{1{\rm gc}} &=& \gamma_{1} \;-\; \iota_{1}\cdot\vb{\omega}_{0} \;+\; \exd \sigma_{1}, \label{eq:ovgamma_1} \\
\Gamma_{2{\rm gc}}  & \equiv & -\;\iota_{2}\cdot\vb{\omega}_{0} - \frac{1}{2}\;\iota_{1}\cdot\left(\vb{\omega}_{1} \;+\frac{}{} \vb{\omega}_{{\rm gc}1} \right) 
+ \exd \sigma_{2},  \label{eq:ovgamma_2} 
\end{eqnarray}
and the extended guiding-center Hamiltonian
\begin{equation}
{\cal H}_{\rm gc} \;\equiv\; {\sf T}_{\rm gc}^{-1}{\cal H} \;=\; {\cal H}_{0{\rm gc}} + \epsilon\,{\cal H}_{1{\rm gc}} + \epsilon^{2}\,{\cal H}_{2{\rm gc}} + \cdots,
\label{eq:ovHam_Lie}
\end{equation}
where 
\begin{eqnarray}
{\cal H}_{1{\rm gc}} &=& {\cal H}_{1} \;-\; {\sf G}_{1}\cdot\exd{\cal H}_{0},
\label{eq:ovH_1} \\
{\cal H}_{2{\rm gc}} &=& -\,{\sf G}_{2}\cdot\exd{\cal H}_{0} \;-\; \frac{1}{2}\;{\sf G}_{1}\cdot\exd\left({\cal H}_{1} + {\cal H}_{1{\rm gc}}\right).
\label{eq:ovHamiltonian_2}
\end{eqnarray}
Here, we use the formulas $\iota_{n}\cdot\vb{\omega} \equiv G_{n}^{\alpha}\omega_{\alpha\beta}\exd Z^{\beta}$ (for an arbitrary two-form $\vb{\omega}$) and ${\sf G}_{n}\cdot\exd{\cal K} \equiv G_{n}^{\alpha}\partial{\cal K}/\partial Z^{\alpha}$ (for an arbitrary scalar field ${\cal K}$), where the summation rule is used. We note that the guiding-center phase-space transformation considered in the present work will contain all terms associated with first-order space-time derivatives of the electric and magnetic fields, which will require us to consider some terms at third order in $\epsilon$ in Eq.~\eqref{eq:ovgamma_Lie}.

In order to construct the extended guiding-center Lagrangian one-form \eqref{eq:ovgamma_Lie}, we will need to evaluate the contractions $\iota_{n}\cdot\vb{\omega}_{0}$ generated by the vector fields 
$({\sf G}_{1}, {\sf G}_{2}, \cdots)$ on the zeroth-order two-form:
\begin{eqnarray} 
\vb{\omega}_{0} \;=\; \exd\gamma_{0} & = & \frac{e}{c} \left( \pd{A_{j}}{x^{i}}\;\exd x^{i} \;+\; \pd{A_{j}}{t}\;\exd t\right) \;\wedge\;\exd x^{j} \;-\; \exd w\wedge \exd t,
\label{eq:omega_0}
\end{eqnarray}
so that we obtain the $n^{\rm th}$-order expression
\begin{equation}
\iota_{n}\cdot\vb{\omega}_{0} = \frac{e}{c}\;{\bf B}\btimes G_{n}^{{\bf x}}\bdot\exd{\bf X} - \left( \frac{e}{c}\,G_{n}^{{\bf x}}\bdot\pd{\bf A}{t} + G_{n}^{w}\right)\;\exd t,
\label{eq:iota_n0}
\end{equation}
where $G_{n}^{{\bf x}}$ and $G_{n}^{w}$ denote the spatial and energy components of the $n$th-order generating vector field ${\sf G}_{n}$. Similarly, we will need to evaluate the contractions $\iota_{n}\cdot\vb{\omega}_{1}$ generated by the vector fields $({\sf G}_{1}, {\sf G}_{2}, \cdots)$ on the first-order two-form $\vb{\omega}_{1} = \exd\gamma_{1}$, which yields the $(n + 1)^{\rm th}$-order expression
\begin{equation}
\iota_{n}\cdot\vb{\omega}_{1} \equiv D_{n}({\bf p}_{0})\bdot\exd{\bf X} - G_{n}^{\bf x}\bdot\left(\pd{{\bf p}_{0}}{t}\;\exd t + \bhat\;\exd p_{\|0} + \pd{{\bf q}_{\bot0}}{J_{0} }\exd J_{0}  + \pd{{\bf q}_{\bot0}}{\theta_{0} }\exd\theta_{0} \right),
\label{eq:iota_n1}
\end{equation}
where ${\bf p}_{0}$ is defined in Eq.~\eqref{eq:p_local} and the spatial components are expressed in terms of the operator \citep{Tronko_Brizard_2015}
\begin{eqnarray}
D_{n}({\bf C}) & \equiv & \left( G_{n}^{p_{\|}}\,\pd{{\bf C}}{p_{\|0}} + G_{n}^{J}\,\pd{{\bf C}}{J_{0}} + G_{n}^{\theta}\,\pd{{\bf C}}{\theta_{0}}\right) \;-\; G_{n}^{\bf x}\btimes\nabla\btimes{\bf C},
\label{eq:Pn_def}
\end{eqnarray}
where ${\bf C}$ is an arbitrary vector function on guiding-center phase space. In what follows, unless it is necessary, we will omit writing the subscript $0$ on 
local particle phase-space coordinates (i.e., $p_{\|0}$ is written as $p_{\|}$).

The purpose of the Lie-transform expressions \eqref{eq:ovgamma_Lie} and \eqref{eq:ovHam_Lie} is to construct a gyroangle-independent extended guiding-center phase-space Lagrangian one-form $\Gamma_{\rm gc}$ and extended guiding-center Hamiltonian 
${\cal H}_{\rm gc}$ in terms of which guiding-center equations are derived. The generic forms considered for the extended guiding-center phase-space Lagrangian one-form \eqref{eq:ovgamma_Lie} and the extended guiding-center Hamiltonian \eqref{eq:ovHam_Lie} in the present work are
\begin{eqnarray}
\Gamma_{\rm gc} &\equiv& \frac{e}{c}\,{\bf A}\bdot\exd{\bf X} \;-\; W\,\exd t \;+\; \epsilon\,{\bf P}_{\rm gc}\bdot\exd{\bf X}  \;+\; \epsilon^{2}J\,\left(\exd\theta - {\bf R}\bdot\exd{\bf X} - {\cal S}\,\exd t\right), \label{eq:Gamma_gc_def} \\
 {\cal H}_{\rm gc} &\equiv& e\,\Phi \;+\; \epsilon\,K_{\rm gc} \;-\; W, \label{eq:Ham_gc_def}
\end{eqnarray}
where the gyroangle-independent guiding-center kinetic momentum ${\bf P}_{\rm gc}$ and guiding-center kinetic energy $K_{\rm gc}$ are expressed as asymptotic series in powers of $\epsilon$, while the vector field ${\bf R} \equiv \nabla\wh{\bot}\bdot\wh{\rho}$ and the scalar field ${\cal S} \equiv (\partial\wh{\bot}/\partial t)\bdot\wh{\rho}$ are required to preserve gyrogauge invariance. See App.~A of the recent paper by \cite{Brizard_2023_gcVM} for an updated discussion on gyrogauge invariance introduced by \cite{RGL_1981,RGL_1983,RGL_1988}. We note that the separation of the guiding-center transformations of the extended guiding-center phase-space Lagrangian one-form \eqref{eq:ovgamma_Lie} and the extended guiding-center Hamiltonian
\eqref{eq:ovHam_Lie} might enable the application of computer algorithms previously used by \cite{Burby_SQ_2013}, but this consideration falls outside the scope of the present work.

\section{\label{sec:Symplectic}Symplectic Polarization Guiding-center Theory}

In the present work, we will consider the {\it symplectic} polarization guiding-center theory, where the guiding-center kinetic momentum ${\bf P}_{\rm gc}$ retains the contribution from the $E\times B$ momentum \eqref{eq:P_E_def}, which will then introduce the polarization drift velocity explicitly in the guiding-center equations of motion. 

In the alternate {\it Hamiltonian} polarization guiding-center theory, on the other hand, the $E\times B$ momentum \eqref{eq:P_E_def} is removed from the guiding-center kinetic momentum and polarization effects enter solely through the guiding-center Hamiltonian. The Hamiltonian case generally requires a different ordering for the electric field (because it produces a polarization displacement that must be compared with the characteristic lowest-order particle gyroradius) and will be considered in a future publication.This dual representation is analogous to the treatment of the perturbed magnetic field in nonlinear gyrokinetic theory \citep{Brizard_Hahm_2007}.

The reader interested in results of the guiding-center Lie-transform perturbation analysis can skip Section \ref{sec:Symplectic} and go to Section \ref{sec:gc_Ham}, where we present the extended guiding-center Hamiltonian structure as well as the (regular) guiding-center Lagrangian, from which guiding-center polarization and magnetization can be derived by functional derivatives with respect to the electric field ${\bf E}$ and magnetic field ${\bf B}$, respectively.

\subsection{\label{sec:first}First-order perturbation analysis}

\subsubsection{First-order symplectic structure}

We begin our perturbation analysis by considering the first-order guiding-center symplectic one-form (\ref{eq:ovgamma_1}), which is now explicitly written in the  {\it symplectic} polarization representation as
\begin{eqnarray}
\Gamma_{1{\rm gc}} & = & \left( P_{\|}\;\bhat \;+\; {\bf P}_{\rm E} \;+\; {\bf q}_{\bot}\right)\bdot\exd{\bf X} \;-\; \frac{e}{c}\;{\bf B}\btimes G_{1}^{{\bf x}}\bdot\exd{\bf X} \;+\; \left( \frac{e}{c}\,G_{1}^{{\bf x}}\bdot\pd{\bf A}{t} \;+\; G_{1}^{w}\right)\;\exd t \nonumber \\
 &=& \left( P_{\|}\,\bhat \;+\; {\bf P}_{\rm E} \right) \bdot\exd{\bf X} \;\equiv\; {\bf P}_{0}\bdot\exd{\bf X},
\label{eq:ovgamma1_Lie}
\end{eqnarray}
where the first-order gauge scalar field $\sigma_{1}$ is not needed and the gyroangle-dependent terms on the right of Eq.~\eqref{eq:ovgamma1_Lie} are eliminated by selecting
\begin{equation}
G_{1}^{{\bf x}} \;=\; {\bf q}_{\bot}\btimes\frac{c\bhat}{eB} \;=\; \frac{1}{m\Omega}\;\pd{{\bf q}_{\bot}}{\theta} \;\equiv\; -\; \vb{\rho}_{0},
\label{eq:G1_x}
\end{equation}
and
\begin{equation}
G_{1}^{w} \;=\; -\,\frac{e}{c}\,G_{1}^{{\bf x}}\bdot\pd{\bf A}{t} \;=\; \frac{e}{c}\,\pd{\bf A}{t}\bdot\vb{\rho}_{0},
\label{eq:G1_w}
\end{equation}
where the gyroangle-dependent vector $\vb{\rho}_{0}$ represents the lowest-order particle gyroradius.  

With $G_{1}^{{\bf x}}$ defined by Eq.~\eqref{eq:G1_x}, the resulting first-order guiding-center phase-space Lagrangian $\Gamma_{1{\rm gc}} = {\bf P}_{0}\bdot\exd{\bf X}$ yields the $(n+1)^{\rm th}$-order contraction 
\begin{equation}
\iota_{n}\cdot\vb{\omega}_{{\rm gc}1} \equiv D_{n}({\bf P}_{0})\bdot\exd{\bf X} - G_{n}^{\bf x}\bdot\left( \bhat\;\exd P_{\|} + \pd{{\bf P}_{0}}{t}\exd t\right),
\label{eq:iotan_1gc}
\end{equation}
where, using the operator \eqref{eq:Pn_def}, the spatial components in Eq.~\eqref{eq:iotan_1gc} are
\begin{eqnarray}
D_{n}({\bf P}_{0}) & = & G_{n}^{p_{\|}}\;\bhat - G_{n}^{\bf x}\btimes\nabla\btimes{\bf P}_{0},
\label{eq:ovP_n_def}
\end{eqnarray}
which contain gyroangle dependent and independent contributions.

\subsubsection{First-order Hamiltonian}

The first-order guiding-center Hamiltonian is determined from Eq.~\eqref{eq:ovH_1} as
\begin{eqnarray}
{\cal H}_{1{\rm gc}}  &=& \mu\,B + \frac{|{\bf P}_{0}|^{2}}{2m} + {\bf P}_{\rm E}\bdot\frac{{\bf q}_{\bot}}{m} - e\,G_{1}^{\bf x}\bdot\nabla\Phi + G_{1}^{w} \nonumber \\
 &=& \mu\,B + \frac{|{\bf P}_{0}|^{2}}{2m} +  {\bf P}_{\rm E}\bdot\frac{{\bf q}_{\bot}}{m} - e\,\vb{\rho}_{0}\bdot{\bf E},
 \label{eq:H1gc_def}
\end{eqnarray}
where the components \eqref{eq:G1_x}-\eqref{eq:G1_w} were substituted, and the electric field is defined as ${\bf E} = -\,\nabla\Phi - c^{-1}\partial{\bf A}/\partial t$. By using the identity
\begin{equation}
{\bf P}_{\rm E}\bdot{\bf q}_{\bot}/m \;=\; {\bf E}\btimes\frac{c\bhat}{B}\bdot{\bf q}_{\bot} \;=\; e\,\vb{\rho}_{0}\bdot{\bf E},
\label{eq:P_rho_id}
\end{equation}
the first-order guiding-center Hamiltonian is automatically gyroangle-independent
\begin{equation}
{\cal H}_{1{\rm gc}} \;=\; \mu\,B + |{\bf P}_{0}|^{2}/(2m) \;=\; \mu B \;+\; P_{\|}^{2}/(2m) \;+\; (m/2)\;\left|{\bf E}\btimes c\bhat/B\right|^{2},
\label{eq:H1_gc}
\end{equation}
which corresponds to the kinetic energy in the frame drifting with the $E\times B$ velocity.

\subsection{\label{sec:second}Second-order perturbation analysis}

\subsubsection{Second-order symplectic structure}

We now proceed with the second-order guiding-center symplectic one-form~(\ref{eq:ovgamma_2}), which is explicitly expressed as
\begin{eqnarray} 
\Gamma_{2{\rm gc}} & = & -\;\left[ \frac{e}{c}{\bf B}\btimes G_{2}^{{\bf x}} \;+\; D_{1}({\bf P}_{2}) \right] \bdot\exd{\bf X} \;+\; \frac{1}{2}\,G_{1}^{{\bf x}}\bdot\left( \pd{{\bf q}_{\bot}}{J}\,\exd J + \pd{{\bf q}_{\bot}}{\theta}\;\exd\theta \right) \nonumber \\
 &&+\; \left( G_{1}^{\bf x}\bdot\pd{{\bf P}_{2}}{t} \;+\; G_{2}^{w} \;+\; \frac{e}{c}\,G_{2}^{\bf x}\bdot\pd{\bf A}{t} \right)\; \exd t \nonumber \\
 & \equiv & \vb{\Pi}_{1}\bdot\exd{\bf X} \;+\; J\;\left(\exd\theta \;-\frac{}{} {\bf R}\bdot\exd{\bf X} \;-\; {\cal S}\;\exd t\right),
\label{eq:ovgamma2_Lie}
\end{eqnarray}
where the second-order gauge scalar field $\sigma_{2}$ is not needed, and, using the definition \eqref{eq:Pn_def} with ${\bf P}_{2} \equiv {\bf P}_{0} + \frac{1}{2}\,{\bf q}_{\bot}$, we find
\begin{eqnarray}
D_{1}({\bf P}_{2}) &=& D_{1}( {\bf P}_{0}) \;+\; \frac{1}{2} \left(g_{1}^{J}\;\pd{{\bf q}_{\bot}}{J} + g_{1}^{\theta}\;\pd{{\bf q}_{\bot}}{\theta} \right) \;+\; J \left[ {\bf R} - \left(\frac{1}{2}\,\tau + \alpha_{1}\right)\bhat \right], \label{eq:D1_P2} 
\end{eqnarray}
where $g_{1}^{J} \equiv G_{1}^{J} - J\,\vb{\rho}_{0}\bdot\nabla\ln B$ and $g_{1}^{\theta} \equiv G_{1}^{\theta} + \vb{\rho}_{0}\bdot{\bf R}$, while $\tau \equiv \bhat\bdot\nabla\btimes\bhat$ and $\alpha_{1} \equiv {\sf a}_{1}:\nabla\bhat$ is defined in terms of the gyroangle-dependent dyadic tensor ${\sf a}_{1} \equiv -\,\frac{1}{2}(\wh{\bot}\wh{\rho} + \wh{\rho}\wh{\bot})$ \citep{Tronko_Brizard_2015}. The first-order guiding-center symplectic momentum $\vb{\Pi}_{1}$ in Eq.~\eqref{eq:ovgamma2_Lie}, which is assumed to be gyroangle-independent, will be determined based on the consistency of the Lie-transform perturbation analysis at the third order (see Sec.~\ref{sec:third}) as well as the guiding-center push-forward derivation of the guiding-center polarization in the absence of a background electric field \citep{Brizard_2013,Tronko_Brizard_2015}.

Substituting these expressions into Eq.~\eqref{eq:ovgamma2_Lie}, we obtain the vector equation
\begin{eqnarray}
\vb{\Pi}_{1} &=& -\,\frac{e}{c}{\bf B}\btimes G_{2}^{\bf x} \;-\; g_{1}^{p_{\|}}\;\bhat \;-\; \vb{\rho}_{0}\btimes\nabla\btimes {\bf P}_{0} \;-\; \frac{1}{2} \left(g_{1}^{J}\;\pd{{\bf q}_{\bot}}{J} + g_{1}^{\theta}\;\pd{{\bf q}_{\bot}}{\theta} \right),
 \label{eq:Pi_1_eq}
 \end{eqnarray}
 where $g_{1}^{p_{\|}} \equiv G_{1}^{p_{\|}} \;-\; J\left(\frac{1}{2}\,\tau \;+\; \alpha_{1}\right)$, while we choose the second-order energy component
 \begin{equation}
 G_{2}^{w} \;=\; \vb{\rho}_{0}\bdot\pd{{\bf P}_{0}}{t} \;-\; \frac{e}{c}\pd{\bf A}{t}\bdot G_{2}^{\bf x},
 \label{eq:G2_w}
 \end{equation}
 where we used 
\begin{equation}
G_{1}^{\bf x}\bdot\pd{{\bf P}_{2}}{t} \;=\; -\;\vb{\rho}_{0}\bdot\pd{{\bf P}_{2}}{t} \;=\; -\;\vb{\rho}_{0}\bdot\pd{{\bf P}_{0}}{t} \;-\; J\,{\cal S}.
\label{eq:J_S}
\end{equation}

Next, from the parallel component of Eq.~\eqref{eq:Pi_1_eq}, we obtain the first-order component
\begin{equation}
G_{1}^{p_{\|}} \;=\; \pd{\vb{\rho}_{0}}{\theta}\bdot\nabla\btimes{\bf P}_{0} \;+\; J \left( \frac{1}{2}\,\tau \;+\; \alpha_{1} \right) \;-\; \Pi_{1\|},
\label{eq:G1_vpar}
\end{equation}
where $\Pi_{1\|} \equiv \bhat\bdot\vb{\Pi}_{1}$ is the parallel component of the first-order symplectic momentum $\vb{\Pi}_{1}$. From the perpendicular components of Eq.~\eqref{eq:Pi_1_eq}, on the other hand, we find
\begin{eqnarray}
G_{2}^{{\bf x}} & = & G_{2\|}^{\bf x}\;\bhat + \left(\frac{\bhat}{m\Omega}\bdot\nabla\btimes{\bf P}_{0}\right)\;\vb{\rho}_{0} + \frac{1}{2} \left( g_{1}^{J}\;\pd{\vb{\rho}_{0}}{J} + g_{1}^{\theta}\; \pd{\vb{\rho}_{0}}{\theta} \right) - \vb{\Pi}_{1}\btimes\frac{\bhat}{m\Omega},
\label{eq:G2_x}
\end{eqnarray}
where $G_{2\|}^{\bf x} \equiv \bhat\bdot G_{2}^{{\bf x}}$. The interpretation of the first-order guiding-center symplectic momentum $\vb{\Pi}_{1}$ will be given in Sec.~\ref{sec:Ham_2}.

\subsubsection{Second-order Hamiltonian}

The second-order guiding-center Hamiltonian is determined from Eq.~\eqref{eq:ovHamiltonian_2} as
\begin{eqnarray}
{\cal H}_{2{\rm gc}}  &=& e\,{\bf E}\bdot \left( G_{2}^{\bf x} \;-\; \frac{1}{2}\,{\sf G}_{1}\cdot\exd\vb{\rho}_{0}\right) \;+\; \frac{e}{2}\;\vb{\rho}_{0}\bdot\nabla{\bf E}\bdot\vb{\rho}_{0} \;-\; \frac{P_{\|}}{m}\,G_{1}^{p_{\|}} \;-\; \Omega\;G_{1}^{J} \nonumber \\
  &&+\; \vb{\rho}_{0}\bdot\left( \mu\nabla B \;+\; \nabla{\bf u}_{\rm E}\bdot m\,{\bf u}_{\rm E} \;+\; \pd{{\bf P}_{0}}{t}\right),
 \label{eq:H2gc_def}
\end{eqnarray}
where $G_{1}^{p_{\|}}$ is given by Eq.~\eqref{eq:G1_vpar}.

Next, we introduce the particle gyroradius 
\begin{eqnarray}
\vb{\rho} \equiv {\bf x} - {\sf T}_{\rm gc}{\bf X} &=& \epsilon\,\vb{\rho}_{0} - \epsilon^{2} \left( G_{2}^{\bf x} - \frac{1}{2}\,{\sf G}_{1}\cdot\exd \vb{\rho}_{0} \right) + \cdots \;=\; \epsilon\,\vb{\rho}_{0} \;+\; \epsilon^{2}\,\vb{\rho}_{1} \;+\; \cdots,
\end{eqnarray}
which is defined as the difference between the particle position ${\bf x}$ and the pull-back ${\sf T}_{\rm gc}{\bf X}$ of the guiding-center position ${\bf X}$, where $\vb{\rho}_{1} \equiv -\,G_{2}^{\bf x} + \frac{1}{2}\,{\sf G}_{1}\cdot\exd\vb{\rho}_{0}$ is the first-order particle gyroradius, where
\begin{eqnarray}
{\sf G}_{1}\cdot\exd\vb{\rho}_{0} &=& g_{1}^{J}\;\pd{\vb{\rho}_{0}}{J} \;+\; g_{1}^{\theta}\; \pd{\vb{\rho}_{0}}{\theta} \;+\; \frac{J}{m\Omega} \left( \nabla\bdot\bhat \;-\; 4\,\alpha_{2}\right)\,\bhat.
 \label{eq:G1_drho}
\end{eqnarray}
Hence, using Eqs.~\eqref{eq:G2_x} and \eqref{eq:G1_drho}, we obtain the first-order particle gyroradius vector
\begin{eqnarray}
\vb{\rho}_{1} &=& \vb{\Pi}_{1}\btimes\frac{\bhat}{m\Omega} - \left[ G_{2\|}^{\bf x} - \frac{J}{m\Omega} \left( \frac{1}{2}\,\nabla\bdot\bhat - 2\,\alpha_{2}\right)\right]\,\bhat \nonumber \\
 &&+ \left(\frac{1}{2}\,\vb{\rho}_{0}\bdot\nabla\ln B - \frac{\bhat}{m\Omega}\bdot\nabla\btimes{\bf P}_{0}\right)\;\vb{\rho}_{0},
 \label{eq:G2_G1d}
\end{eqnarray}
where $4\,\alpha_{2} \equiv -\,\partial\alpha_{1}/\partial\theta$.

We now note that the gyroangle-dependent part $\wt{G}_{1}^{J} \equiv G_{1}^{J} - \langle G_{1}^{J}\rangle$ can be defined such that the right side of Eq.~\eqref{eq:H2gc_def} only contains terms that are gyroangle-independent. Hence, we find
\begin{eqnarray}
\Omega\,\wt{G}_{1}^{J} &=& -\,\frac{p_{\|}}{m}\;\wt{G}_{1}^{p_{\|}} \;-\; e\,\wt{\vb{\rho}}_{1}\bdot{\bf E} \;-\; \frac{2J}{m\Omega}\;e\,{\sf a}_{2}:\nabla{\bf E} \nonumber \\
 &&+\; \vb{\rho}_{0}\bdot\left(  \mu\nabla B \;+\; \nabla{\bf P}_{\rm E}\bdot{\bf u}_{\rm E} \;+\; \pd{{\bf P}_{0}}{t}\right),
 \label{eq:G1J_tilde}
\end{eqnarray}
where ${\sf a}_{2} \equiv \frac{1}{4}\;(\wh{\bot}\,\wh{\bot} - \wh{\rho}\,\wh{\rho}) = -\,\frac{1}{4}\partial{\sf a}_{1}/\partial\theta$ and the gyroangle-dependent part of Eq.~\eqref{eq:G1_vpar} is
\begin{equation}
\wt{G}_{1}^{p_{\|}} \;\equiv\; G_{1}^{p_{\|}} \;-\; \langle G_{1}^{p_{\|}}\rangle \;=\;  \pd{\vb{\rho}_{0}}{\theta}\bdot\nabla\btimes{\bf P}_{0} \;+\; J\;\alpha_{1}.
\end{equation}
The second-order guiding-center Hamiltonian is, thus, defined as
\begin{eqnarray}
{\cal H}_{2{\rm gc}} &=& \frac{P_{\|}}{m}\;\left(\Pi_{1\|} - \frac{1}{2}\,J\,\tau \right) \;-\; \Omega\,\langle G_{1}^{J}\rangle \;+\; \frac{J\,c}{2B}\;\left(\mathbb{I} - \bhat\bhat\right):\nabla{\bf E} \;-\; \langle\vb{\rho}_{1}\rangle\bdot e\,{\bf E} 
 \label{eq:H2gc} \\
 &=& \frac{P_{\|}}{m}\;\left(\Pi_{1\|} - \frac{1}{2}\,J\,\tau \right) - \Omega\,\langle G_{1}^{J}\rangle + \nabla\bdot\left(\frac{e}{2}\,\langle\vb{\rho}_{0}\vb{\rho}_{0}\rangle\bdot{\bf E}\right) + \left(\vb{\Pi}_{1} - \frac{1}{2}\,J\,\nabla\btimes\bhat\right)
 \bdot{\bf u}_{\rm E},
\nonumber
\end{eqnarray}
where $\langle G_{1}^{p_{\|}}\rangle = \frac{1}{2}\,J\,\tau - \Pi_{1\|}$ and the gyroangle-averaged first-order particle gyroradius is
\begin{eqnarray}
 \langle\vb{\rho}_{1}\rangle &=& \frac{J}{2\,m\Omega}\left[\left(\mathbb{I} - \bhat\bhat\right)\vb{\cdot}\nabla\ln B + \left(\nabla\vb{\cdot}\bhat\right)\bhat \right] + \vb{\Pi}_{1}\vb{\times}\frac{\bhat}{m\Omega} \nonumber \\
  &\equiv& -\nabla\vb{\cdot}\left( \frac{1}{2}\langle\vb{\rho}_{0}\vb{\rho}_{0}\rangle\right) + \left( \vb{\Pi}_{1} - \frac{J}{2}\nabla\vb{\times}\bhat\right)\vb{\times}\frac{\bhat}{m\Omega}
 \end{eqnarray}
 where $\langle\vb{\rho}_{0}\vb{\rho}_{0}\rangle = (J/m\Omega)\;(\mathbb{I} - \bhat\bhat)$. Hence, we now need expressions for $\langle G_{1}^{J}\rangle$ and $\vb{\Pi}_{1}$ in order to obtain an explicit expression for the second-order guiding-center Hamiltonian \eqref{eq:H2gc}, which are determined at third order in our perturbation analysis.

\subsection{\label{sec:third}Third-order perturbation analysis}

The third-order term in the guiding-center phase-space Lagrangian one-form \eqref{eq:ovgamma_Lie} is expressed as
\begin{eqnarray}
\Gamma_{3{\rm gc}} & = & -\iota_{3}\cdot\vb{\omega}_{0} \;-\; \iota_{2}\cdot\vb{\omega}_{{\rm gc}1} \;+\; \frac{\iota_{1}}{3}\cdot\exd\left( \iota_{1}\cdot\vb{\omega}_{1} + \frac{\iota_{1}}{2}\cdot\vb{\omega}_{{\rm gc}1} \right) \;+\; \exd \sigma_{3}.
\label{eq:ovgamma_3}
\end{eqnarray}
In what follows, the gauge function $\sigma_{3}$ will play an important role in completing the guiding-center phase-space transformation, while the third-order guiding-center Hamiltonian will not be needed in the present guiding-center formulation.

\subsubsection{Third-order symplectic structure}

The remaining components $(G_{2\|}^{\bf x}, \langle G_{1}^{J}\rangle, G_{1}^{\theta})$ and the first-order guiding-center momentum $\vb{\Pi}_{1}$ will now be determined from the momentum components of the third-order guiding-center symplectic one-form \eqref{eq:ovgamma_3}
\begin{eqnarray} 
\Gamma_{3{\bf p}} & \equiv & \left[ G_{2\|}^{\bf x} \;+\; \pd{D_{1}({\bf P}_{3})}{p_{\|}}\bdot\vb{\rho}_{0} \;+\; \pd{\sigma_{3}}{p_{\|}} \right]\;\exd 
p_{\|} \;+\; \left[ \frac{2}{3}\,G_{1}^{\theta} + \pd{D_{1}({\bf P}_{3})}{J}\bdot\vb{\rho}_{0} + \pd{\sigma_{3}}{J} \right]\;\exd J \nonumber \\
 &  &+\; \left[ -\,\frac{2}{3}\,G_{1}^{J} + \pd{D_{1}({\bf P}_{3})}{\theta}\bdot\vb{\rho}_{0} + \pd{\sigma_{3}}{\theta} \right]\;\exd \theta ,
\label{eq:Gamma_3_p}
\end{eqnarray}
where  ${\bf P}_{3} \equiv \frac{1}{2}\,{\bf P}_{0} + \frac{1}{3}\,{\bf q}_{\bot}$, so that
\begin{eqnarray}
D_{1}({\bf P}_{3}) &=& \frac{1}{2}\;G_{1}^{p_{\|}}\,\bhat \;+\; \frac{1}{3} \left(G_{1}^{J}\;\pd{{\bf q}_{\bot}}{J} + G_{1}^{\theta}\;
\pd{{\bf q}_{\bot}}{\theta}\right) \;+\; \vb{\rho}_{0}\btimes\nabla\btimes{\bf P}_{3}. \label{eq:D1_3rd}
\end{eqnarray}

Since $\partial\vb{\rho}_{0}/\partial p_{\|} = 0$, the $p_{\|}$-component of Eq.~\eqref{eq:Gamma_3_p} suggests that we define the new gauge function
\begin{equation}
\ov{\sigma}_{3} \;\equiv\; \sigma_{3} \;+\; D_{1}({\bf P}_{3})\bdot\vb{\rho}_{0} \;=\; \sigma_{3}
\;-\; \frac{2}{3}\;J\,G_{1}^{\theta},
\label{eq:sigma3_ov}
\end{equation}
where the last expression follows from Eq.~\eqref{eq:D1_3rd}. Using the new gauge function \eqref{eq:sigma3_ov}, the momentum components \eqref{eq:Gamma_3_p}, therefore, become
\begin{eqnarray}
\Gamma_{3{\bf p}} & = & \left( G_{2\|}^{\bf x} \;+\; \pd{\ov{\sigma}_{3}}{p_{\|}} \right)\;\exd p_{\|} \;+\; \left( G_{1}^{\theta} \;+\; 
\pd{\ov{\sigma}_{3}}{J} \right)\;\exd J \nonumber \\
 &  &+\; \left( \pd{\ov{\ov{\sigma}}_{3}}{\theta} \;-\; G_{1}^{J} \;+\; \frac{J\bhat}{m\Omega}\bdot\nabla\btimes{\bf P}_{0}\right)\;\exd \theta,
\label{eq:ovgamma_3gc_final}
\end{eqnarray}
where, using Eq.~\eqref{eq:D1_3rd}, we introduced the identities
\begin{eqnarray*}
D_{1}({\bf P}_{3})\bdot\pd{\vb{\rho}_{0}}{J} & \equiv & -\;\frac{1}{3}\,G_{1}^{\theta}, \\
D_{1}({\bf P}_{3})\bdot\pd{\vb{\rho}_{0}}{\theta} & \equiv & \frac{1}{3}\,G_{1}^{J} \;+\; \frac{2J\bhat}{m\Omega}\bdot\nabla\btimes{\bf P}_{3},
\end{eqnarray*}
so that we can also introduce yet another gauge function
\begin{equation}
\ov{\ov{\sigma}}_{3} \;\equiv\; \ov{\sigma}_{3} \;-\; \frac{1}{3} \left( 2\,J\;\vb{\rho}_{0}\bdot{\bf R} \;+\; J\;\pd{\vb{\rho}_{0}}{\theta}\bdot\nabla\ln B\right)
\label{eq:sigma3_ov_ov}
\end{equation}
in the $\theta$-component of Eq.~\eqref{eq:Gamma_3_p}. By requiring that the momentum components \eqref{eq:ovgamma_3gc_final} vanish, we now obtain the definitions
\begin{eqnarray}
G_{1}^{J} & \equiv & -\;\frac{J\bhat}{m\Omega}\bdot\nabla\btimes{\bf P}_{0} \;+\; \pd{\ov{\ov{\sigma}}_{3}}{\theta}, \label{eq:G1_J_eq} \\
G_{2\|}^{\bf x} & \equiv & -\;\pd{\ov{\sigma}_{3}}{p_{\|}}, \label{eq:G2_xpar_eq} \\
G_{1}^{\theta} & \equiv & -\;\pd{\ov{\sigma}_{3}}{J}. \label{eq:G1_theta_eq}
\end{eqnarray}
Hence, the components $G_{2\|}^{\bf x}$ and $G_{1}^{\theta}$ are determined from the third-order gauge function $\ov{\sigma}_{3}$, which is determined from Eq.~\eqref{eq:sigma3_ov_ov}, while $\ov{\ov{\sigma}}_{3}$ is determined from the gyroangle-dependent part $\wt{G}_{1}^{J} \equiv \partial\ov{\ov{\sigma}}_{3}/\partial\theta$ obtained from Eq.~\eqref{eq:G1J_tilde}. Since these gyroangle-dependent components are not needed in what follows, however, they will not be derived here.

\subsubsection{\label{sec:Ham_2}Second-order guiding-center Hamiltonian}

From Eq.~\eqref{eq:G1_J_eq}, we immediately conclude that $\langle G_{1}^{J}\rangle$ must be defined as
\begin{equation}
\langle G_{1}^{J}\rangle \;\equiv\; -\;\frac{J\bhat}{m\Omega}\bdot\nabla\btimes{\bf P}_{0},
\label{eq:Banos_J}
\end{equation} 
which was obtained in previous derivations \citep{Brizard_1995,Madsen_2010,Frei_2020}, so that the second-order guiding-center Hamiltonian \eqref{eq:H2gc} becomes
\begin{equation}
{\cal H}_{2{\rm gc}} \;=\; -\;\nabla\bdot\left(\frac{e}{2}\,\langle\vb{\rho}_{0}\vb{\rho}_{0}\rangle\bdot{\bf E}\right) \;+\; \left(\vb{\Pi}_{1} + \frac{1}{2}\,J\,\nabla\btimes\bhat\right) \bdot\frac{{\bf P}_{0}}{m},
 \label{eq:H2gc_Pi0}
 \end{equation}
 where we used the identity
\begin{equation}
\frac{J}{2m\Omega}\,\left(\mathbb{I} - \bhat\bhat\right)\bdot e\,{\bf E} \;\equiv\;  \frac{e}{2}\,\langle\vb{\rho}_{0}\vb{\rho}_{0}\rangle\bdot{\bf E} \;\equiv\; \frac{1}{2}\,J\,\bhat\btimes{\bf u}_{\rm E}.
\label{eq:gc_quad}
\end{equation}
In previous works \citep{RGL_1981,Hahm_1996,Miyato_Scott_2009,Madsen_2010,Frei_2020}, the choice for the first-order symplectic momentum $\vb{\Pi}_{1} = -\frac{1}{2}\,J\tau\,\bhat$ was used to eliminate the Ba\~{n}os drift \citep{Banos_1967,Northrop_Rome_1978} from the guiding-center velocity (i.e., $\partial{\cal H}_{2{\rm gc}}/\partial P_{\|} = 0$), which is instead added to the definition of the guiding-center parallel momentum \eqref{eq:G1_vpar}. This choice, therefore, yields the second-order guiding-center Hamiltonian ${\cal H}_{2{\rm gc}} = (J\bhat/2)\bdot\nabla\btimes{\bf u}_{\rm E}$. 

A different choice adopted by \cite{Tronko_Brizard_2015} for the first-order symplectic momentum
\begin{equation}
\vb{\Pi}_{1{\rm pol}} \;\equiv\; -\,\frac{1}{2}\,J\;\nabla\btimes\bhat,
\label{eq:Pi1_def}
\end{equation}
on the other hand, was previously shown to yield an exact Lie-transform perturbation derivation of the standard guiding-center polarization derived by \cite{Kaufman_1986} in the absence of an electric field. While this choice still eliminates the Ba\~{n}os drift from the guiding-center velocity, it also yields an expression for the second-order guiding-center Hamiltonian \eqref{eq:H2gc_Pi0} that exactly represents the guiding-center finite-Larmor-radius (FLR) correction to the electrostatic potential energy $e\,\Phi$ \citep{Brizard_2023}:
\begin{equation}
{\cal H}_{2{\rm gc}} \;=\; -\;\nabla\bdot\left( \frac{e}{2}\,\langle\vb{\rho}_{0}\vb{\rho}_{0}\rangle\bdot{\bf E} \right).
\label{eq:Hgc_2}
\end{equation}
Finally, we note that recent numerical studies of particle and guiding-center orbits in axisymmetric magnetic fields \citep{Brizard_Hodgeman_2023} have shown that guiding-center orbits are faithful (i.e., remain close) to particle orbits only if second-order effects, including the correction \eqref{eq:Pi1_def}, are included, which confirms earlier results by \cite{Belova_2003}.

\section{\label{sec:gc_Ham}Guiding-center Hamiltonian Dynamics} 

In this Section, we summarize the results of the Lie-transform perturbation analysis of the guiding-center Lagrangian dynamics presented in Sec.~\ref{sec:Symplectic}, and we remove the explicit $\epsilon$ scaling by restoring the physical mass $\epsilon\,m \rightarrow m$. Hence, we find the guiding-center phase-space extended one-form
\begin{eqnarray}
\Gamma_{\rm gc} &=& \left(\frac{e}{c}\;{\bf A} \;+\; \vb{\Pi}_{\rm gc}\right)\bdot\exd{\bf X} \;+\; J\;\left(\exd\theta \;-\frac{}{} {\bf R}\bdot\exd{\bf X} \;-\; {\cal S}\,\exd t\right) \;-\; W\,\exd t,
\label{eq:Gamma_gc_primitive}
\end{eqnarray}
where the guiding-center symplectic momentum
\begin{equation}
\vb{\Pi}_{\rm gc} \;=\; P_{\|}\,\bhat \;+\; {\bf P}_{\rm E} \;-\; \frac{J}{2}\,\nabla\btimes\bhat
\label{eq:Pi_gc_def}
\end{equation}
includes the higher-order polarization correction \eqref{eq:Pi1_def}. The extended guiding-center Hamiltonian, on the other hand, is expressed as
\begin{equation}
{\cal H}_{\rm gc} \;=\; e\,\Phi \;+\; K_{\rm gc}  \;-\; W,
\label{eq:Phi_star}
\end{equation}
 where the guiding-center kinetic energy in the drifting frame is
 \begin{equation}
 K_{\rm gc} \;=\; \mu\,B \;+\; \frac{P_{\|}^{2}}{2m} \;+\; \frac{m}{2}\,|{\bf u}_{\rm E}|^{2} \;-\; \nabla\bdot\left( \frac{J\bhat}{2}\btimes{\bf u}_{\rm E}\right),
 \label{eq:K_gc_def}
  \end{equation}
which includes the FLR correction \eqref{eq:Hgc_2} to the electrostatic potential energy $e\,\Phi$. We note that the presence of the gyrogauge fields $({\cal S},{\bf R})$ in Eq.~\eqref{eq:Gamma_gc_primitive} guarantees gyrogauge invariance of the guiding-center equations of motion derived from them.

\subsection{Extended guiding-center Poisson bracket}

The extended guiding-center Poisson bracket $\{\;,\; \}_{\rm gc}$ is obtained by, first, constructing an 8$\times$8 matrix out of the components of the extended guiding-center Lagrange two-form $\vb{\omega}_{\rm gc} = \exd\Gamma_{\rm gc}$ and, then, invert this matrix to obtain the extended guiding-center Poisson matrix, whose components are the fundamental brackets $\{Z^{\alpha}, Z^{\beta}\}_{\rm gc}$. From these components, we obtain the  extended guiding-center Poisson bracket
\begin{eqnarray}
\{{\cal F}, {\cal G}\}_{\rm gc} &=& \left(\pd{\cal F}{W}\,\frac{\partial^{*}{\cal G}}{\partial t} - \frac{\partial^{*}{\cal G}}{\partial t}\,\pd{\cal G}{W}\right) \;+\; \frac{{\bf B}^{*}}{B_{\|}^{*}}\bdot\left(\nabla^{*}{\cal F}\,\pd{\cal G}{P_{\|}} - \pd{\cal F}{P_{\|}}\,\nabla^{*}{\cal G}\right) \nonumber \\
  &&-\; \frac{c\bhat}{eB_{\|}^{*}}\bdot\nabla^{*}{\cal F}\btimes\nabla^{*}{\cal G} \;+\; \left(\pd{\cal F}{\theta}\,\pd{\cal G}{J} - \pd{\cal F}{J}\,\pd{\cal G}{\theta}\right), 
   \label{eq:gcPB_ext}
 \end{eqnarray}
 where 
 \begin{equation}
 \frac{e}{c}\,{\bf B}^{*} \;=\; \frac{e}{c}\,{\bf B} \;+\; \nabla\btimes\vb{\Pi}_{\rm gc} \;-\; J\,\nabla\btimes{\bf R}, \label{eq:B_star}
 \end{equation}
and the guiding-center Jacobian is ${\cal J}_{\rm gc} = (e/c)\,B_{\|}^{*} \equiv (e/c)\,\bhat\bdot{\bf B}^{*}$. In addition, we have introduced the definitions
 \begin{eqnarray}
 \frac{\partial^{*}}{\partial t} &\equiv& \pd{}{t} \;+\; {\cal S}\;\pd{}{\theta}, \label{eq:t_star} \\
 \nabla^{*} &\equiv& \nabla \;+\; {\bf R}^{*}\;\pd{}{\theta} \;-\; \left( \frac{e}{c}\pd{{\bf A}^{*}}{t} + J\,\nabla{\cal S}\right)\pd{}{W}, \label{eq:grad_star}
 \end{eqnarray}
where ${\bf R}^{*} \equiv {\bf R} + \frac{1}{2}\,\nabla\btimes\bhat$. We note that the Poisson bracket \eqref{eq:gcPB_ext} can be expressed in divergence form as
\begin{equation}
\{{\cal F}, {\cal G}\}_{\rm gc} \;=\; \frac{1}{B_{\|}^{*}}\pd{}{Z^{\alpha}}\left(B_{\|}^{*}\frac{}{}{\cal F}\;\left\{ Z^{\alpha},\; {\cal G}\right\}_{\rm gc}\right),
\label{eq:gcPB_div}
\end{equation}
and that it automatically satisfies the Jacobi identity
 \begin{equation}
 \left\{{\cal F},\frac{}{}\{{\cal G},{\cal K}\}_{\rm gc}\right\}_{\rm gc} +  \left\{{\cal G},\frac{}{}\{{\cal K},{\cal F}\}_{\rm gc}\right\}_{\rm gc} +  \left\{{\cal K},\frac{}{}\{{\cal F},{\cal G}\}_{\rm gc}\right\}_{\rm gc} = 0
 \end{equation}
 since the extended guiding-center Lagrange two-form $\vb{\omega}_{\rm gc} = \exd\Gamma_{\rm gc}$ is exact (i.e., $\exd\vb{\omega}_{\rm gc} = \exd^{2}\Gamma_{\rm gc} = 0$) provided $\nabla\bdot{\bf B}^{*} = 0$. 
 
 Next, we note that the operators \eqref{eq:t_star} and \eqref{eq:grad_star} contain the gyrogauge-invariant combinations $\partial/\partial t + {\cal S}\,\partial/\partial\theta$ and $\nabla + {\bf R}\;\partial/\partial\theta$, while Eqs.~\eqref{eq:B_star} and \eqref{eq:grad_star} include the gyrogauge-independent vector fields \citep{Ye_Kaufman_1992,Brizard_2023_gcVM}
 \begin{equation}
\left( \nabla\btimes{\bf R},  \nabla{\cal S} \;-\; \pd{\bf R}{t} \right) \;=\; \left( -\,\frac{1}{2}\,\epsilon_{ijk}\,b^{i}\;\nabla b^{j}\btimes\nabla b^{k}, -\,\nabla\bhat\btimes\bhat\bdot\pd{\bhat}{t} \right),
\label{eq:RS_vector}
\end{equation}
 where $\epsilon_{ijk}$ denotes the completely-antisymmetric Levi-Civita tensor.

\subsection{Guiding-center Hamilton equations}

 The guiding-center Hamilton equations $\dot{Z}^{\alpha} \equiv \{ Z^{\alpha}, {\cal H}_{\rm gc}\}_{\rm gc}$ include the guiding-center velocity
 \begin{equation}
 \dot{\bf X} \;=\; \frac{P_{\|}}{m}\;\frac{{\bf B}^{*}}{B_{\|}^{*}} \;+\; {\bf E}^{*}\btimes\frac{c\bhat}{B_{\|}^{*}}, 
 \label{eq:Xgc_dot}
 \end{equation}
 where $\bhat\bdot\dot{\bf X} = P_{\|}/m$ defines the parallel guiding-center velocity, and the guiding-center parallel force
 \begin{equation}
 \dot{P}_{\|} \;=\; e\,{\bf E}^{*}\bdot\frac{{\bf B}^{*}}{B_{\|}^{*}},
 \label{eq:Pgc_dot}
 \end{equation}
 where the modified electric field is represented as
 \begin{eqnarray}
 e\,{\bf E}^{*} &=& e\,{\bf E} - \pd{\vb{\Pi}_{\rm gc}}{t} - \nabla K_{\rm gc} + J \left(\pd{\bf R}{t} - \nabla{\cal S}\right), \label{eq:E_star}
  \end{eqnarray}
 and the gyroangle angular velocity
 \begin{eqnarray}
 \dot{\theta} &\equiv& \pd{K_{\rm gc}}{J} \;+\; {\cal S} \;+\; \dot{\bf X}\bdot{\bf R}^{*} = \Omega \;-\; \nabla\bdot\left(\frac{\bhat}{2}\btimes{\bf u}_{\rm E}\right) \;+\; {\cal S} \;+\; \dot{\bf X}\bdot{\bf R}^{*}. 
 \label{eq:thetagc_dot}
  \end{eqnarray}

We note that the reduced guiding-center equations of motion \eqref{eq:Xgc_dot}-\eqref{eq:Pgc_dot} satisfy the guiding-center Liouville equation
 \begin{equation}
 \pd{B_{\|}^{*}}{t} \;=\; -\;\nabla\bdot\left(B_{\|}^{*}\;\dot{\bf X}\right) \;-\; \pd{}{P_{\|}}\left(B_{\|}^{*}\;\dot{P}_{\|}\right),
 \label{eq:gc_Liouville}
 \end{equation} 
 where
 \begin{eqnarray*} 
 \nabla\bdot\left(B_{\|}^{*}\;\dot{\bf X}\right) &=& \nabla\btimes{\bf E}^{*}\bdot c\bhat - e\,{\bf E}^{*}\bdot\frac{c}{e}\nabla\btimes\bhat = -\,\bhat\bdot\pd{{\bf B}^{*}}{t} - e\,{\bf E}^{*}\bdot\pd{{\bf B}^{*}}{P_{\|}}
 \end{eqnarray*}
 and
 \begin{eqnarray*}
 \pd{}{P_{\|}}\left(B_{\|}^{*}\;\dot{P}_{\|}\right) &=& e\,{\bf E}^{*}\bdot\pd{{\bf B}^{*}}{P_{\|}} + {\bf B}^{*}\bdot e\pd{{\bf E}^{*}}{P_{\|}} = e\,{\bf E}^{*}\bdot\pd{{\bf B}^{*}}{P_{\|}} - {\bf B}^{*}\bdot \pd{\bhat}{t},
 \end{eqnarray*}
 where we made use of the modified Faraday's law $\partial{\bf B}^{*}/\partial t = -c\,\nabla\btimes{\bf E}^{*}$.
 
\subsection{Eulerian variations of the guiding-center Lagrangian}
 
The results of the Lie-transform perturbation analysis carried out in Section \ref{sec:third} can also be used to construct the following (regular) guiding-center Lagrangian
 \begin{eqnarray}
 L_{\rm gc} &=& \left(\frac{e}{c}\,{\bf A} + \vb{\Pi}_{\rm gc} - J\,{\bf R}\right)\bdot\dot{\bf X} \;+\; J\,\dot{\theta} \;-\; \left(e\,\Phi \;+\frac{}{} K_{\rm gc} \;+\; J\,{\cal S}\right) \nonumber \\
  &\equiv& (e/c)\,{\bf A}^{*}\bdot\dot{\bf X} \;+\; J\,\dot{\theta} - H_{\rm gc},
 \label{eq:Lag_gc}
 \end{eqnarray}
 where $\vb{\Pi}_{\rm gc}$ and $K_{\rm gc}$ are defined in Eqs.~\eqref{eq:Pi_gc_def} and \eqref{eq:K_gc_def}, respectively. The guiding-center Euler-Lagrange equations are derived from this Lagrangian as
 \begin{equation}
 \dot{P}_{\|}\,\bhat \;=\; e\,{\bf E}^{*} \;+\; (e/c)\,\dot{\bf X}\btimes{\bf B}^{*}, 
 \end{equation}
 \begin{equation}
 \bhat\bdot\dot{\bf X} \;=\; P_{\|}/m, 
 \end{equation}
which are associated with virtual displacements $\delta{\bf X}$ and $\delta P_{\|}$, respectively. From these equations, we easily recover the guiding-center Hamilton equations \eqref{eq:Xgc_dot} and \eqref{eq:Pgc_dot}. Likewise, the guiding-center Euler-Lagrange equation associated with the virtual displacement $\delta J$ yields Eq.~\eqref{eq:thetagc_dot}, while the virtual displacement $\delta\theta$ yields $\dot{J} = 0$ as a result of Noether's Theorem.

In addition to variations with respect to guiding-center phase-space coordinates, the guiding-center Lagrangian \eqref{eq:Lag_gc} can also be varied with respect to the electric and magnetic fields $(\delta{\bf E}, \delta{\bf B})$, which respectively yield expressions for the guiding-center polarization and magnetization \citep{Brizard_2008}. Here, the Eulerian variation of the guiding-center Lagrangian \eqref{eq:Lag_gc} is expressed as \citep{Brizard_2023_gcVM}
 \begin{eqnarray}
\delta L_{\rm gc} &\equiv& \left( \frac{e}{c}\delta{\bf A} \;+\; \delta\vb{\Pi}_{\rm gc} \right)\bdot\dot{\bf X} \;-\; \left(e\,\delta\Phi \;+\frac{}{} \delta K_{\rm gc} \right) \;-\; J \left(\delta{\cal S} \;+\frac{}{} \dot{\bf X}\bdot\delta{\bf R}\right) \nonumber \\
 &=& \left( \frac{e}{c}\delta{\bf A}\bdot\dot{\bf X} - e\,\delta\Phi\right) + \vb{\pi}_{\rm gc}\bdot\delta{\bf E} + \left( \vb{\mu}_{\rm gc} + \vb{\pi}_{\rm gc}\btimes\frac{{\bf P}_{0}}{mc}\right)\bdot\delta{\bf B} \;+\; ({\rm FLR}),
  \label{eq:delta_Psi}
  \end{eqnarray}
 where the FLR corrections, which are ignored in Eq.~\eqref{eq:delta_Psi}, are calculated to first order in a recent paper \citep{Brizard_2023_gcVM}. Here, the guiding-center electric dipole moment $\vb{\pi}_{\rm gc}$ is defined as \citep{Pfirsch_1984,Pfirsch_Morrison_1985}
\begin{equation}
\vb{\pi}_{\rm gc} \;\equiv\; \frac{e\bhat}{\Omega}\btimes\left(\dot{\bf X} \;-\frac{}{} {\bf u}_{\rm E}\right),
\label{eq:pi_gc}
\end{equation}
while the guiding-center magnetic dipole moment $\vb{\mu}_{\rm gc} + \vb{\pi}_{\rm gc}\btimes{\bf P}_{0}/(mc)$ is defined as the sum of the intrinsic magnetic dipole moment
\begin{equation}
\vb{\mu}_{\rm gc} \;\equiv\; \mu \left(-\,\bhat \;+\; \frac{1}{\Omega}\,\frac{d\bhat}{dt}\btimes\bhat\right),
\label{eq:mu_gc}
\end{equation}
which includes the gyrogauge correction associated with $\delta{\cal S} + \dot{\bf X}\bdot\delta{\bf R}$ \citep{Brizard_2023_gcVM}, and the moving electric-dipole contribution \citep{Jackson_1975}, expressed in terms of the lowest-order guiding-center momentum ${\bf P}_{0} = P_{\|}\,\bhat + {\bf P}_{\rm E}$. By ignoring these FLR corrections, the guiding-center polarization and magnetization are defined as moments of the guiding-center electric and magnetic dipole moments
\begin{eqnarray}
\vb{\cal P}_{\rm gc} &=& \int_{P} F_{\rm gc}\;\vb{\pi}_{\rm gc}, \label{eq:Pol_gc} \\
\vb{\cal M}_{\rm gc} &=& \int_{P} F_{\rm gc}\;\left( \vb{\mu}_{\rm gc} + \vb{\pi}_{\rm gc}\btimes\frac{{\bf P}_{0}}{mc}\right), \label{eq:Mag_gc}
\end{eqnarray}
where the guiding-center phase-space density $F_{\rm gc} \equiv {\cal J}_{\rm gc}\,F$ includes the guiding-center Jacobian ${\cal J}_{\rm gc}$ and the notation $\int_{P}$ includes an integral over guiding-center momentum space as well as a sum over particle species. 

Finally, we note that, in the absence of an electric field, the classical expression \citep{Kaufman_1986} for the guiding-center electric dipole moment $\vb{\pi}_{\rm gc} = (e\bhat/\Omega)\btimes\dot{\bf X}$ is derived by Lie-transform perturbation method \citep{Tronko_Brizard_2015} only if the first-order polarization correction \eqref{eq:Pi1_def} is used.

\section{Summary}

In the present work, a set of higher-order guiding-center Hamilton equations was derived by Lie-transform perturbation method for the case of time-dependent electric and magnetic fields that satisfy the standard guiding-center space-time orderings. The second-order corrections in the guiding-center Hamiltonian represented finite-Larmor-radius corrections of the lowest-order electrostatic potential energy $e\,\Phi$. Additional second-order corrections in the guiding-center Lagrangian \eqref{eq:Lag_gc} included gyrogauge-invariance contributions to the guiding-center Hamiltonian and Poisson bracket as well as corrections leading to the correct guiding-center polarization.

When we turn our attention to the self-consistent interactions of the charged-particle guiding-centers and the electromagnetic fields associated with plasma confinement, we need to derive a set of higher-order guiding-center Vlasov-Maxwell equations. Work presented elsewhere \citep{Brizard_2023_gcVM} considered the variational formulation of the higher-order guiding-center Vlasov-Maxwell equations derived from the guiding-center Lagrangian \eqref{eq:Lag_gc} and the Maxwell Lagrangian density. 

According to this variational principle, using the guiding-center Liouville equation \eqref{eq:gc_Liouville}, the guiding-center Vlasov equation for the guiding-center phase-space density $F_{\rm gc} \equiv {\cal J}_{\rm gc}\,F$ is written in divergence form as
\begin{equation}
\pd{F_{\rm gc}}{t} \;+\; \nabla\bdot\left(\dot{\bf X}\frac{}{} F_{\rm gc}\right) \;+\; \pd{}{P_{\|}} \left(\dot{P}_{\|}\frac{}{} F_{\rm gc}\right) \;=\; 0,
\label{eq:gc_Vlasov}
\end{equation}
while the Maxwell equations with particle sources are expressed as
\begin{eqnarray}
\nabla\bdot{\bf E} &=& 4\pi\,\varrho \;\equiv\; 4\pi \left(\varrho_{\rm gc} \;-\frac{}{} \nabla\bdot\vb{\cal P}_{\rm gc}\right), \label{eq:div_E} \\
\nabla\btimes{\bf B} - \frac{1}{c}\,\pd{\bf E}{t} &=& \frac{4\pi}{c}\,{\bf J} \;\equiv\;  \frac{4\pi}{c}\left( {\bf J}_{\rm gc} \;+\; \pd{\vb{\cal P}_{\rm gc}}{t} \;+\; c\,\nabla\btimes\vb{\cal M}_{\rm gc}\right), \label{eq:curl_B}
\end{eqnarray}
where the guiding-center charge and current densities are $\varrho_{\rm gc} = \int_{P} e\,F_{\rm gc}$ and ${\bf J}_{\rm gc} = \int_{P} e\,\dot{\bf X}\;F_{\rm gc}$. Here, the guiding-center polarization charge density $\varrho_{\rm pol} \equiv -\nabla\bdot
\vb{\cal P}_{\rm gc}$ and current density ${\bf J}_{\rm pol} \equiv \partial\vb{\cal P}_{\rm gc}/\partial t$ are derived from the guiding-center polarization \eqref{eq:Pol_gc}, while the guiding-center magnetization current density ${\bf J}_{\rm mag} \equiv c\,\nabla\btimes\vb{\cal M}_{\rm gc}$ is derived from the guiding-center magnetization \eqref{eq:Mag_gc}. The remaining source-free Maxwell equations are Faraday's law $\partial{\bf B}/\partial t = -c\,\nabla\btimes{\bf E}$ and Gauss's law $\nabla\bdot{\bf B} = 0$. We note that the guiding-center Vlasov-Maxwell variational principle also guarantees the existence of exact energy-momentum conservation laws, derived by the Noether method \citep{Brizard_2008}. Our recent work \citep{Brizard_2023_gcVM} has considered the set of higher-order guiding-center Vlasov-Maxwell equations obtained by explicitly imposing the quasineutrality constraint $\varrho_{\rm gc} = \nabla\bdot\vb{\cal P}_{\rm gc}$.
 
Future work will explore the Hamiltonian structure of the guiding-center Vlasov-Maxwell equations, when guiding-center polarization and magnetization are included, which will generalize our previous work \citep{Brizard_2021_gcVM}, and its development is motivated by the desire to construct structure-preserving numerical algorithms \citep{Morrison_2017} using an important set of reduced Vlasov-Maxwell equations.
 
\vspace*{0.1in}

\noindent
{\bf Acknowledgments}

\vspace*{0.1in}
The Author wishes to thank the referees for their insightful comments, which led to the final manuscript. The Author would also like to dedicate this paper to the memory of Allan N. Kaufman (1927-2022), whose pioneering efforts in introducing Lie-transform perturbation methods in plasma physics \citep{Kaufman_1978} led to the landmark papers by \cite{Cary_1981} and \cite{Cary_1983}, culminating with the review papers by \cite{Brizard_Hahm_2007} and \cite{Cary_Brizard_2009}. 

\vspace*{0.1in}

\noindent
{\bf Funding}

\vspace*{0.1in}
The present work was supported by the National Science Foundation grant PHY-2206302.

\vspace*{0.1in}

\noindent
{\bf Declaration of Interests}

\vspace*{0.1in}
The Author reports no conflict of interest.

\bibliographystyle{jpp}

\bibliography{gc_extended}

\end{document}